# DESIGN OF A COMPACT REVERSIBLE READ-ONLY-MEMORY WITH MOS TRANSISTORS

Sadia Nowrin, Papiya Nazneen and Lafifa Jamal

Department of Computer Science and Engineering, University of Dhaka, Bangladesh

*ABSTRACT*

*Energy conservative devices are the need of the modern technology which leads to the development of reversible logic. The synthesis of reversible logic has become an intensely studied area as it overcomes the problem of power dissipation associated with irreversibility. Storage device such as Read-Only-Memory (ROM) can be realized in a reversible way with low power dissipation. The reversibility of ROM has not been yet realized in literature and hence, this paper presents a novel reversible ROM with its Complementary Metal Oxide Semiconductor (CMOS) realization. On the way to present the architecture of reversible ROM, we propose a new reversible gate named as Nowrin Papiya (NP) gate. All the proposed circuits and gates are realized with CMOS based pass transistor logic. Finally, an algorithm as well as several theorems on the numbers of gates, transistors and garbage outputs have been presented to show the optimality of the reversible ROM. Simulations using Microwind DSCH software has been shown to verify the correctness of the proposed design. The comparative results prove that the proposed designs are efficient and optimized in terms of numbers of gates, transistors, garbage outputs, quantum cost and delay.*

## 1. INTRODUCTION

Power dissipation that leads to overheating is becoming one of the major issues in today's world. The amount of energy dissipated holds a direct relationship to the number of bits erased during computation. According to R. Landauer's research [1], traditional computer dissipates *KT*ln2 joules of energy for each bit of information loss, where *K* is the Boltzmann constant and *T* is the operating temperature. Bennett [2] stated that *KT*ln2 Joules of energy would not be dissipated if the computation is carried out in a reversible way. That gave rise to reversible logic circuit which does not erase information and input vector can be uniquely reconstructed from the output vector.

In recent years, reversible logic has gained attention in low power CMOS design [2], quantum computing, nanotechnology, quantum dot cellular automata, cryptography etc. Zhirnov et al. [3] proved that power dissipation in any future CMOS will lead to an impossible heat removal problem. The number of transistors will double every 18 months according to Gordon Moore [4]. However, if Zhirnov and Moore's Law will continue to be in effect, speeding-up of CMOS devices will be impossible at some point in 2020 or earlier.

In this work, we present the design methodologies of a compact reversible ROM with its CMOS realization. To the best of our knowledge, this is the first work to realize reversible ROM with MOS transistors. We propose different components of reversible ROM with their transistor design and finally, we propose the reversible architecture of ROM. All the proposed circuits are optimized in terms of gate count, transistor count, garbage output, quantum cost and delay.

## 2. LITERATURE REVIEW

This section presents the basic ideas and theories of reversible logic with some popular reversible gates which are relevant to this research work.

### 2.1 Reversible Gate

A reversible logic gate is an *n*-input, *n*-output circuit having a unique output pattern [5] for each possible input pattern. There is a one-to-one mapping between inputs and outputs. Thus, an $n \times n$ reversible logic gate can be represented as $I_v \leftrightarrow O_v$, where, $I_v = (I_1, I_2, I_3, \ldots, I_{n-1}, I_n)$ is the input vector and $O_v = (O_1, O_2, O_3, \ldots, O_{n-1}, O_n)$ is the output vector.

### 2.2 Optimization Issues

The main challenge of designing a reversible circuit is to optimize different cost metrics or parameters as follows:

#### 2.2.1 Number of Gates

The total number of gates used in a circuit. A reversible circuit must use minimum possible number of gates.

#### 2.2.2 Constant Inputs

The inputs that are added to an $n \times k$ function to make it reversible are called constant inputs. The number of constant inputs should be minimum to design reversible logic circuits.

#### 2.2.3 Garbage Outputs

Unwanted or unused outputs of a reversible gate or circuit required to maintain reversibility are known as garbage outputs [5].

#### 2.2.4 Quantum Cost

One of the most important challenges in reversible logic design is to optimize quantum cost. The quantum cost can be derived by substituting the reversible logic gates of a circuit by a cascade of elementary quantum gates [6]. Elementary quantum gates realize quantum circuits that are inherently reversible and manipulate qubits rather than pure logic values [7]. The state of a qubit for two pure logic states can be expressed like $|\Psi\rangle = \alpha |0\rangle + \beta |1\rangle$, where $|0\rangle$ and $|1\rangle$ denote 0 and 1, respectively, and $\alpha$ and $\beta$ are complex numbers such that $|\alpha|^2 + |\beta|^2 = 1$. The most used elementary quantum gates are the NOT gate (a single qubit is inverted), the controlled-NOT (CNOT) gate (the target qubit is inverted if the single control qubit is 1), the controlled-V gate (also known as a square root of NOT, since two consecutive V operations are equivalent to an inversion), and the controlled V+ gate (which performs the inverse operation of the V gate and thus is also a square root of NOT) [7].

### 2.2.5 Delay

The delay of a logic circuit is the maximum number of gates in a path from any input line to the output line [8]. According to [9], [10], the delay of any 3× 3 reversible gate can be computed by calculating its logical depth when it is designed from smaller 1×1 and 2× 2 reversible gates with unit delay. It is denoted by ∆. For example, the Fredkin gate requires 5∆ delay [9]. In this research work, we consider the logical depth as a measure of delay.

### 2.3 Popular Reversible Logic Gates

In this section, we present the basic reversible logic gates which are related to our research work.

### 2.3.1 Feynman Gate

Feynman gate [11] is a 2× 2 reversible gate that maps two inputs (A, B) to two outputs (P = A, Q = A⊕ B). Since this is a 2×2 gate, the quantum cost is 1 [11].

### 2.3.2 Fredkin Gate

Fredkin gate [12] is a 3×3 reversible gate that maps three inputs (A, B, C) to three outputs (P = A, Q = A′B⊕ AC, R = A′C⊕ AB). The quantum cost of Fredkin gate is 5 [12].

### 2.3.3 Toffoli Gate

Toffoli gate [13] is a 3×3 reversible gate with the mapping of (*A*, *B*, *C*) to (*P = A, Q = B, R = AB* ⊕ *C*), where *A*, *B*, *C* are the inputs and *P*, *Q*, *R* are the outputs. The quantum cost of Toffoli gate is 5 [13].

### 2.3.4 HL Gate

HL gate [14] is a 4×4 reversible gate that maps four inputs (*A*, *B*, *C*, *D*) to four outputs (*P = AB′* ⊕ *B′C* ⊕ *BD′*, *Q = AB* ⊕ *B′C* ⊕ *BD*, *R = A′B* ⊕ *B′C* ⊕ *BD* and *S = AB′* ⊕ *BC* ⊕ *B′D*). The quantum cost of HL gate is 7 [14].

## 3. BACKGROUND STUDY

A ROM that can store $2^n$ words each of *m* bit is referred to as $2^n \times m$ bit ROM, where *n* is the number of address inputs and *m* is the number of data bits. A $2^n \times m$ ROM requires one *k*-to-$2^k$ row decoder, one *(n-k)*-to-$2^{(n-k)}$ column decoder, 2n registers each consists of *m* number of D FFs to store data inputs and finally the output buffer to store the data outputs. There is no existing design of reversible ROM in literature. But, the components (Reversible Decoder and D FF) have been realized by many researchers. If the components are designed in an optimized way, the final design of reversible ROM will also be optimized. Various transistor design of reversible decoder have been proposed in [15,16]. Shamsujjoha and Babu [15] proposed the transistor design of a fault tolerant reversible decoder which requires a large number of transistors. Ravish et al. [16] only mentioned the transistor count of a reversible decoder without showing the transistor design and it requires more transistors than [16]. Different approaches of reversible D FF have been proposed in [5,8,10,17–19]. Mamun et al. [17] proposed a reversible D FF which shows no design for the complement output $Q'$. Existing reversible D FFs in [5,8,10,18] requires large number of gates, quantum cost and delay. Thapliyal and Vinod [19] proposed the transistor design of a reversible D latch. But the [19] design for the complement output requires more quantum cost than other existing designs.

## 4. PROPOSED TRANSISTOR REALIZATION OF REVERSIBLE GATES

In this section, we propose the transistor implementation of reversible logic gates that are required to design the components of reversible ROM.

### 4.1. Proposed Transistor Realization of Feynman Gate

The transistor implementation of reversible Feynman gate is shown in Figure 1(a). When control input $A$ is 0, transistor 1 is in on state and input $B$ is passed straight to output $Q$. When control input $A$ is 1, transistor 2 and transistor 3 are in on state and input $B$ is inverted to pass to output $Q$. This design requires only 3 transistors and achieves 62.5% reduction over the existing designs [19] and [20] in terms of transistor count. Our proposed implementation is completely reversible in nature as it is suitable for both forward and backward computation.

Forward Computation:

$$P = A,$$
$$\text{If } A = 0 \text{ then } Q = B$$
$$\text{Else } Q = B'$$

Backward Computation:

$$A = P,$$
$$\text{If } P = 0 \text{ then } B = Q$$
$$\text{Else } B = Q'$$

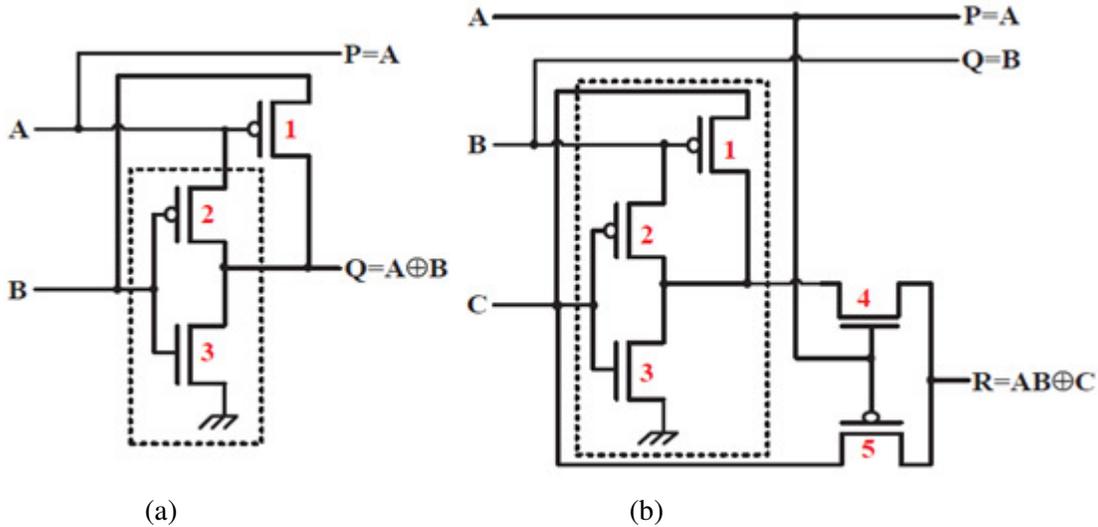

Figure1. Transistor Realization of (a) Feynman Gate and (b) Toffoli Gate

## 4.2. Proposed Transistor Realization of Toffoli Gate

The transistor implementation of Toffoli gate is shown in Figure 1(b). Transistors 1, 2 and 3 in the dotted block represents the transistors in Feynman gate shown in Figure 1(a) and performs EX-OR operation of input $B$ and $C$. When control input $A$ is 1, transistor 4 is in on state and ($B \oplus C$) is passed to output $R$. When control input $A$ is 0, transistor 5 is in on state and input $C$ is passed to output $R$. This design requires only 5 transistors and achieves 58.3% and 68.75% reduction in terms of transistor count over the existing designs [19] and [20], respectively. The proposed implementation is completely reversible in nature. The forward and backward calculation can be explained as follows:

Forward Computation:
$$P = A, Q = B,$$
$$\text{If } A = 1 \text{ and } B = 1 \text{ then } R = C'$$
$$\text{Else } R = C$$

Backward Computation:
$$A = P, B = Q,$$
$$\text{If } P = 1 \text{ and } Q = 1 \text{ then } C = R'$$
$$\text{Else } C = R$$

## 4.3. Proposed Transistor Realization of Hasan Lafifa (HL) Gate

We realize existing HL Gate [14] by 13 transistors as shown in Figure 2. Input $A$ and $D$ are EX-ORed in first dotted block using transistors 1, 2 and 3 which represents three transistors of Feynman gate shown in Figure 1(a). When control input $B$ is 0, transistor 4 is in on state and ($A \oplus D$) is passed to output $S$. Here, input $B$ works as the control input to derive output $Q$ and $S$. When control input $B$ is 1, transistor 5 is in on state and input $C$ is passed to output $S$. Input $B$ controls the transistors 6 and 7 as the transistors 4 and 5 to produce output $Q$. Transistors 8, 9, 10

73

and 11, 12, 13 in the last two dotted blocks performs $(B \oplus Q)$ to produce output $R$ and $(A \oplus R)$ to produce output $P$, respectively. The proposed transistor implementation is completely reversible as computation works in both forward and backward direction.

Forward Computation:

$$\text{If } B = 0 \text{ then } P = A \oplus C, Q = R = C \text{ and } S = A \oplus D$$
$$\text{Else } P = D', Q = A \oplus D, R = A' \oplus D \text{ and } S = C$$

Backward Computation:

$$\text{If } R = 0 \text{ then } A = P \text{ and } D = P \oplus Q'S \oplus Q$$
$$\text{Else } A = P' \text{ and } D = P' \oplus Q'S \oplus Q$$
$$\text{If } Q = 0 \text{ then } B = R \text{ and } C = Q \oplus S$$
$$\text{Else } B = R' \text{ and } C = Q'S \oplus Q$$

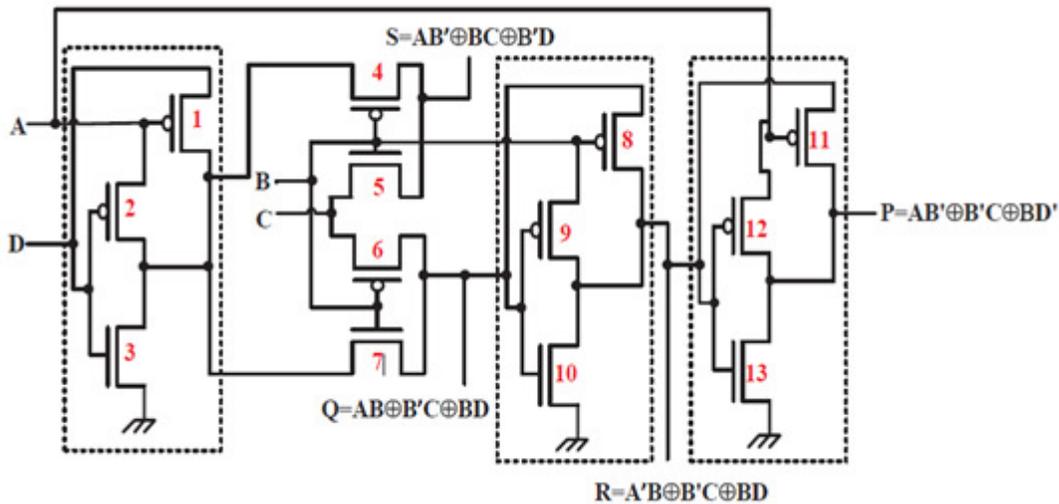

Figure 2 : Transistor Design of Reversible HL Gate

## 5. PROPOSED DESIGN OF REVERSIBLE ROM

The major components of a 2n ×m ROM are decoder and single bit memory cell such as D Flip-Flop. In this section, we present the designs of these components in a reversible way to realize the whole architecture of a reversible ROM.

### 5.1 Proposed Transistor Realization of Reversible Decoder

L.Jamal et al [14] proposed a reversible 2-to-4 decoder using HL gate which requires 13 transistors according to Figure 2. The existing design of reversible decoder [14] is also optimized in terms of garbage output and quantum cost. The performance comparison on transistor count between proposed and existing 2-to-4 reversible decoders [15], [16] is shown in Table I.

Table 1 : Comparison of Reversible 2-To-4 Decoders

| Circuit | Transistor Count |
|---|---|
| **Proposed Circuit** | 13 |
| **Existing Circuit [15]** | 20 |
| **Existing Circuit [16]** | 24 |
| **Improvement(%)w.r.t [15]** | 35 |
| **Improvement(%)w.r.t [16]** | 45.83 |

A 3-to-8 reversible decoder can be realized by applying four outputs of a 2-to-4 reversible decoder to four separate Fredkin gates. So, a generalized $n$-to-$2^n$ reversible decoder needs one ($k$-1)-to-$2^{k-1}$ reversible decoder with additional $2^{k-1}$ number of Fredkin gates.

***Theorem 1:*** If $T$ is the total number of transistors to design an $n$-to-$2^n$ reversible decoder, then $T = 4 \times 2^n - 3$.

***Proof:*** We prove the above statement by induction:

According to our design, a 2-to-4 reversible decoder requires 13 (= $4 \times 2^2 - 3$) transistors. So, the statement holds for base $n = 2$.

Assume for $n = k$, a $k$-to-$2^k$ reversible decoder can be realized by $4 \times 2^k - 3$ no. of transistors.

Now, a ($k$+1)-to-$2^{k+1}$ decoder contains a $k$-to-$2^k$ decoder with $2^k$ number of Fredkin gates. Each Fredkin gate requires 4 transistors [19] resulting in total:

$$T = (4 \times 2^k - 3) + 4 \times 2^k$$
$$= 4 \times 2^{k+1} - 3$$

Thereby, the statement holds for $n = k + 1$.

So, If $T$ is the total number of transistors to design an $n$-to-$2^n$ reversible decoder, then $T = 4 \times 2^n - 3$.

### 5.2 Proposed Reversible 4× 4 NP gate

In this section, a new 4×4 reversible gate named NP gate is proposed which maps input vector $I$ ($A, B, C, D$) to output vector $O$ ($P, Q, R, S$). The output is defined by: $P = A$, $Q = A'B \oplus AC'$, $R = A'C \oplus AB$, $S = A'C \oplus AB \oplus D$. Proposed NP gate requires quantum cost of 5. The block diagram and quantum circuit of NP gate are shown in Figures 3(a) and 3(b) respectively. We can verify from the corresponding truth table shown in Table II that 16 input and 16 output combinations of the proposed NP gate have one-to-one mapping between them. So, NP gate satisfies the condition of reversibility.

<em>International Journal of VLSI design & Communication Systems (VLSICS) Vol.6, No.5, October 2015</em>

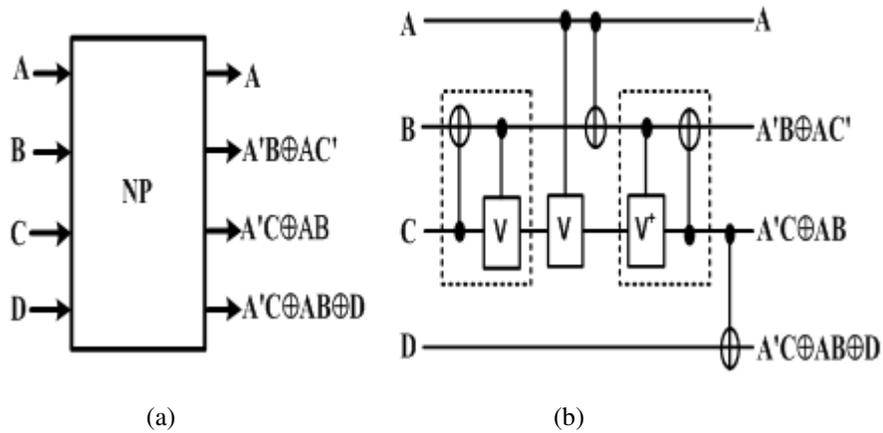

(a)                          (b)

Figure3. (a) Block Diagram and (b) Quantum Circuit of NP gate

Table 2: Truth Table of NP Gate

| Input | | | | Output | | | |
|---|---|---|---|---|---|---|---|
| A | B | C | D | P | Q | R | S |
| 0 | 0 | 0 | 0 | 0 | 0 | 0 | 0 |
| 0 | 0 | 0 | 1 | 0 | 0 | 0 | 1 |
| 0 | 0 | 1 | 0 | 0 | 0 | 1 | 1 |
| 0 | 0 | 1 | 1 | 0 | 0 | 1 | 0 |
| 0 | 1 | 0 | 0 | 0 | 1 | 0 | 0 |
| 0 | 1 | 0 | 1 | 0 | 1 | 0 | 1 |
| 0 | 1 | 1 | 0 | 0 | 1 | 1 | 1 |
| 0 | 1 | 1 | 1 | 0 | 1 | 1 | 0 |
| 1 | 0 | 0 | 0 | 1 | 0 | 0 | 0 |
| 1 | 0 | 0 | 1 | 1 | 0 | 0 | 1 |
| 1 | 0 | 1 | 0 | 1 | 0 | 1 | 1 |
| 1 | 0 | 1 | 1 | 1 | 0 | 1 | 0 |
| 1 | 1 | 0 | 0 | 1 | 1 | 0 | 0 |
| 1 | 1 | 0 | 1 | 1 | 1 | 0 | 1 |
| 1 | 1 | 1 | 0 | 1 | 1 | 1 | 1 |
| 1 | 1 | 1 | 1 | 1 | 1 | 1 | 0 |

## 5.3. Transistor Realization of NP Gate

NP gate can be realized by 9 transistors as shown in Figure 4. Transistors 1 and 2 in the first dotted block inverts input *C*. Input *A* controls transistors 3 and 4 to pass either *C'* or *B* to output *Q*. Similarly input *A* controls transistors 5 and 6 to pass either input *B* or *C* to output *R*. Transistors 7, 8 and 9 performs $R \oplus D$ to produce output *S*. The proposed transistor implementation is completely reversible as computation works in both forward and backward direction.

Forward Computation:

$$P = A,$$
If $A = 0$ then $Q = B$ and $R = C$ and $S = C \oplus D$
Else $Q = C'$ and $R = B$ and $S = B \oplus D$

Backward Computation:

$$A = P,$$
If $P = 0$ then $B = Q$ and $C = R$ and $D = R \oplus S$
Else $C = Q'$ and $B = R$ and $D = R \oplus S$

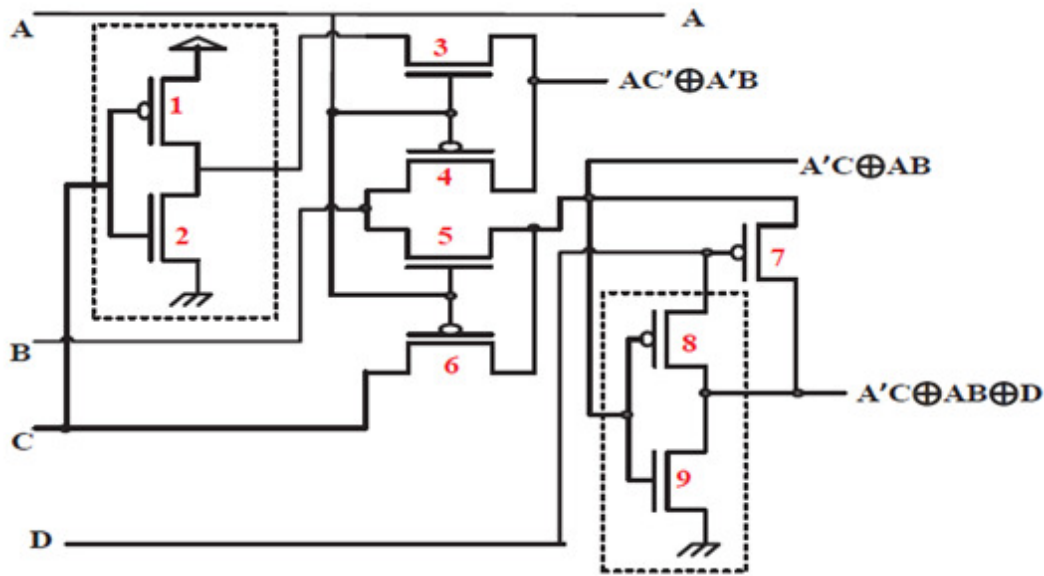

Figure 4. Transistor Implementation of NP Gate

## 5.4. Proposed Design of Reversible D Flip-Flop

The characteristic equation of D FF is written as $Q = E'Q \oplus DE$ which can be constructed using the proposed NP gate as shown in Figure 5(a). Figure 4 shows that an NP gate requires 9 transistors. So, our proposed reversible D FF requires only 9 transistors to produce output $Q$. The ratio of improvements of our proposed design with respect to the existing approaches [10], [17] and [19] in terms of gate count, garbage output, quantum cost and delay are shown in Table 3. According to this table, our proposed design requires only 1 gate, 5 quantum cost, 5Δ delay, 9 transistors and produces 2 garbage outputs. Proposed Reversible D FF (a) with Output $Q$ and (b) with Output $Q$ and $Q'$. The complement output $Q'$ can be produced by adding a Feynman gate as shown in Figure 5(b). Feynman gate requires total 3 transistors as shown in Figure 1(a). So, our proposed reversible D FF for complement output needs additional three transistors resulting in total 12 transistors. Since proposed NP gate has a delay of 5Δ and Feynman gate has a delay of 1Δ, our proposed reversible D FF for complement output requires total delay of 6Δ. Table 4 compares our proposed reversible D FF for complement output with existing designs [18], [10], [17] and [19]. Though the transistor count is not mentioned in existing approaches, but other cost parameters such as garbage output, quantum cost and delay are much optimized in our proposed approach than existing designs in literature.

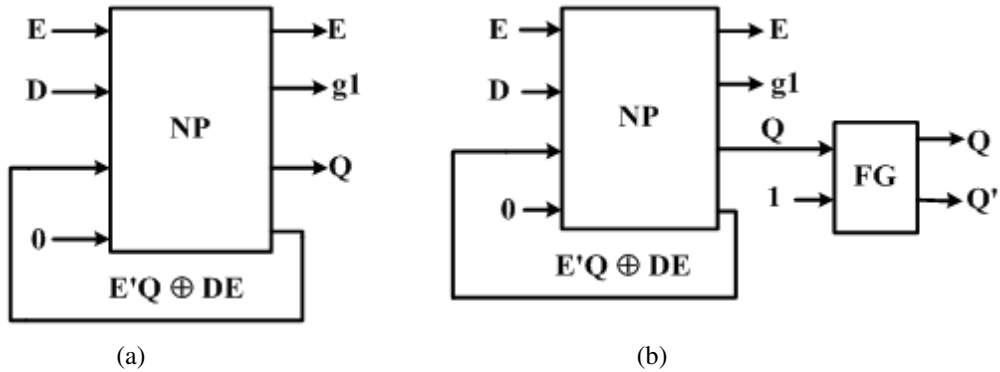

Figure5. Proposed Reversible D FF (a) with Output $Q$ and (b) with Output $Q$ and $Q'$

Proposed reversible DFFs for output $Q$ and $Q'$ are simulated using Microwind DSCH-3.5 [21] software on a computer, which has Pentium (R) Dual-Core CPU with 2.50 GHz clock speed and 2 GB RAM. The timing diagrams of the simulation of proposed reversible DFFs for output $Q$ and $Q'$ are shown in Figure 6.

Table 3 : Comparison of Reversible D FFS with Q Output

| Circuit | GC | GO | QC | TC | DL |
| --- | --- | --- | --- | --- | --- |
| Proposed Circuit | 1 | 2 | 5 | 9 | 5Δ |
| Existing Circuit [5] | 1 | 2 | 6 | - | 6Δ |
| Existing Circuit [10] | 2 | 2 | 6 | - | 6Δ |
| Existing Circuit [17] | 2 | 2 | 6 | - | 6Δ |
| Existing Circuit [19] | 2 | 2 | 6 | 12 | 6Δ |
| Improvement(%)w.r.t [5] | 0 | 0 | 16.67 | - | 16.67 |
| Improvement(%)w.r.t [10] | 50 | 0 | 16.67 | - | 16.67 |
| Improvement(%)w.r.t [17] | 50 | 0 | 16.67 | - | 16.67 |
| Improvement(%)w.r.t [19] | 50 | 0 | 16.67 | 25 | 16.67 |

Where, GC=Gate Count, GO=Garbage Output, QC=Quantum Cost, TC=Transistor Count, DL=Delay

Table 4: Comparison of Reversible D FFS with Q and Q' Output

| Circuit | GC | GO | QC | TC | DL |
|---|---|---|---|---|---|
| Proposed Circuit | 2 | 1 | 6 | 12 | 6Δ |
| Existing Circuit [18] | 2 | 2 | 7 | - | 7Δ |
| Existing Circuit [5] | 2 | 2 | 7 | - | 7Δ |
| Existing Circuit [10] | 3 | 2 | 7 | - | 7Δ |
| Existing Circuit [8] | 2 | 2 | 9 | - | 9Δ |
| Improvement(%)w.r.t [18] | 0 | 50 | 14.28 | - | 14.28 |
| Improvement(%)w.r.t [5] | 0 | 50 | 14.28 | - | 14.28 |
| Improvement(%)w.r.t [10] | 0 | 0 | 14.28 | - | 14.28 |
| Improvement(%)w.r.t [8] | 0 | 50 | 33.33 | - | 14.28 |

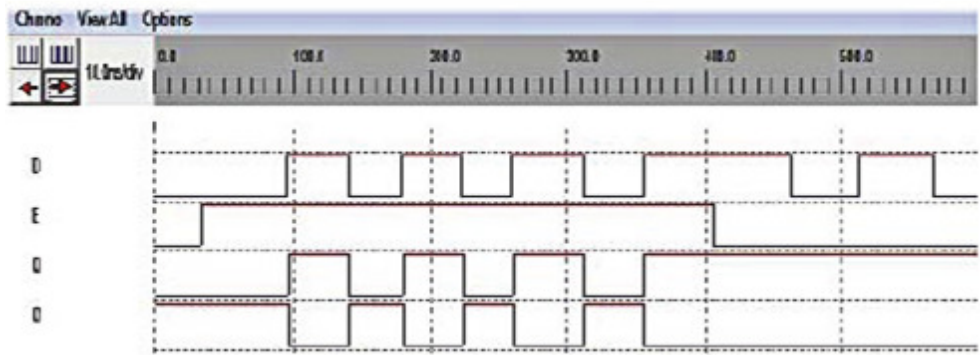

Figure6. Simulation of the Proposed Reversible D FF with Output $Q$ and $Q'$

## 5.5. Proposed Design of Reversible ROM

This section presents the reversible architecture of a $2n \times m$ ROM as shown in Figure 8. In this design, a $k$-to-$2^k$ row decoder selects one row of multiple words, where a $(n-k)$-to-$2^{(n-k)}$ column decoder selects the appropriate word of that row. This selection is performed by Toffoli gate. The D FFs of the selected register pass the data bits to Fredkin gate which works as the output buffer for the final outputs $Q_0$ to $Q_m$. The whole design can be realized by CMOS transistor if each block is replaced by its corresponding transistor design proposed in this paper. The proposed design is simulated using Microwind DSCH-3.5 [21] software on a computer, which has Pentium (R) Dual-Core CPU with 2.50 GHz clock speed and 2 GB RAM. Figure 9 shows the timing diagram of a small 4 × 2 reversible ROM using DSCH-3.5 [21] which verifies that our proposed circuit works correctly. Here, the data inputs ($R_1D_0$ to $R_4D_1$) of the four registers are passed to the outputs ($R_1Q_0$ to $R_4Q_1$) based on the address inputs $I_1$ and $I_2$.

---
**Algorithm 1. $2^n \times m$ Reversible ROM**

---

1. Take one $k$-to-$2^k$ reversible row decoder and one $(n-k)$-to-$2^{n-k}$ column decoder
2. for each row $r = 1$ to $2^k$ in ROM do
3.    for each column $c = 1$ to $2^{n-k}$ do
4.       Take one Toffoli gate $T_{r,c}$
5.          if $c=1$ then
6.             $T_{r,c}[I_1]$ = rth output of row decoder
7.          else
8.             $T_{r,c}[I_1] = T_{r,(c-1)}[O_1]$
9.     end if
10.         if $r = 1$ then
11.     $T_{r,c}[I_2]$ = rth output of column decoder
12.     else
13.       $T_{r,c}[I_2] = T_{(r-1),c}[O_2]$
14.     end if
15.     $T_i[I_3] = 0$
16.     for $j = 1 \to m$ do
17.       Take one reversible D FF $FF_j$
18.       $FF_j[D] = D_j$, primary data input of the ROM
19.       if $c=1$ then
20.          $FF_j[E] = T_j[O_3]$
21.       else
22.          $FF_j[E] = E$ output $FF_j - 1$
23.       end if
24.     end for
25.   end for
26. end for

---

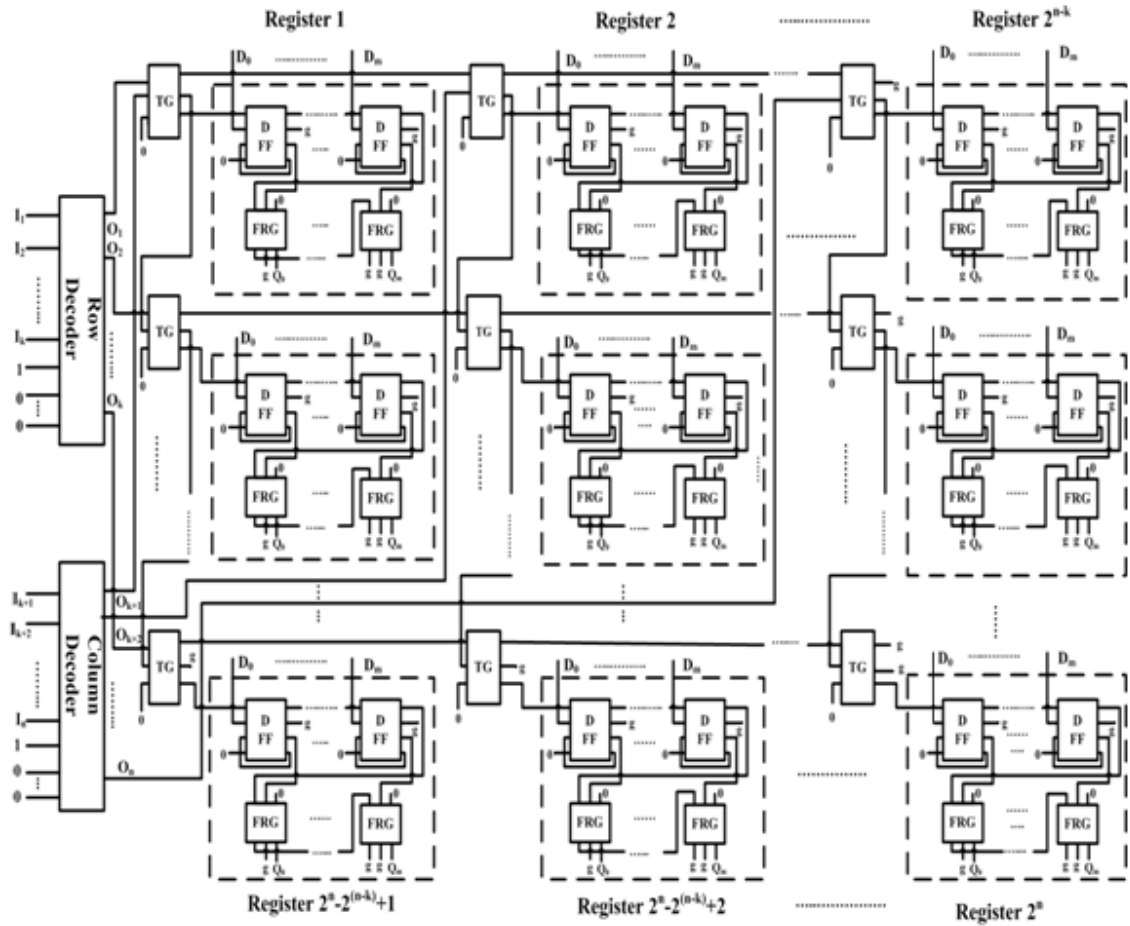

Figure 7. Proposed Design of Reversible $2^n \times m$ ROM

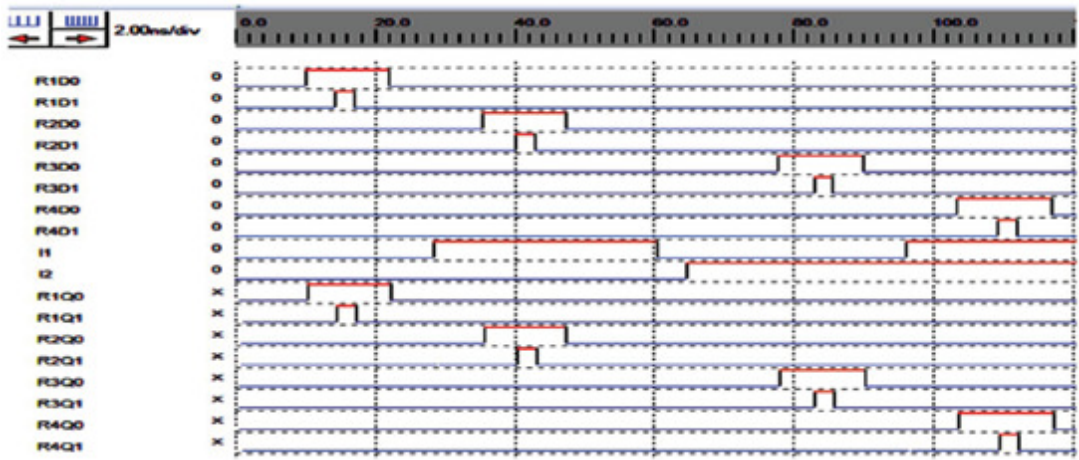

Figure 8. Timing Diagram of Reversible $4 \times 2$ ROM

**Theorem 2:** A $2^n \times m$ Reversible ROM can be realized by $2^k + 2^n(2^{-k} + 2m + 1) - 6$ gates where $n$ be the number of address input and $m$ be the data bit.

**Proof:** A $2^n \times m$ reversible ROM contains one $k$-to-$2^k$ row decoder and one $(n-k)$-to-$2^{n-k}$ column decoder. According to theorem presented in [14], total gates needed in decoder part is:

$$2^k - 3 + 2^{n-k} - 3$$
$$= 2^k + 2^{n-k} - 6.$$

Each of the $2^n$ registers consists of $m$ number of D FFs and each D FF needs one NP gate resulting in total $m \times 2^n$ D FFs. In addition, The ROM requires $2^n$ Toffoli gates and $m \times 2^n$ Fredkin gates for AND and COPY operation. Therefore, total gate count is:

$$2^k + 2^{n-k} - 6 + (m \times 2^n) + 2^n + (m \times 2^n)$$
$$= 2^k + 2^n(2^{-k} + 2m + 1) - 6$$

**Theorem 3:** A $2^n \times m$ Reversible ROM generates $n + 2^k + 2^n(2m + 2^{-k} + 1) - 4$ garbage outputs where $n$ is the number of address input and $m$ is the data bit.

**Proof:** According to [14], Garbage outputs produced from decoder part is $k - 2 + n - k - 2 = n - 4$.

According to our design, Toffoli gates from the last row and last column generates one garbage except the last one producing two garbages which results in total $2^k - 1 + 2^{n-k} - 1 + 2 = 2^k + 2^{n-k}$ garbage outputs. Each of the $m \times 2^n$ D FFs generates one garbage bit.

Again, among $(2^n \times m)$ Fredkin gates, $2^n$ Fredkin gates produces two garbages and rest of the $(2^n \times m - 2^n)$ produce one garbage resulting in total $2 \times 2^n + 2^n \times m - 2^n = 2^n(m+1)$ garbages.

So, total garbages produced from a reversible ROM is:

$$n - 4 + m \times 2^n + 2^k + 2^{n-k} + 2^n(m + 1)$$
$$= n + 2^k + 2^n(2m + 2^{-k} + 1) - 4$$

**Theorem 4:** If $T$ is the total transistor count for a $2^n \times m$ Reversible ROM, then $T = 2^n(4 \times 2^{-k} + 13m + 5) + 4 \times 2^k - 6$.

**Proof:** According to Theorem 1, the number of transistors in decoder part :
$(4 \times 2^k - 3 + 4 \times 2^{n-k} - 3) = 4 \times 2^k + 4 \times 2^{n-k} - 6$.

Each of the $(m \times 2^n)$ D FFs requires 9 transistors resulting in total $9(m \times 2^n)$ transistors.

Again, there are $2^n$ Toffoli gates each having 5 transistors and $(2^n \times m)$ Fredkin gates each having 4 transistors resulting in total $2^n(5 + 4m)$ transistors. So,

$$T = 4 \times 2^k + 4 \times 2^{n-k} - 6 + 9(m \times 2^n) + 2^n(5 + 4m)$$
$$= 2^n(4 \times 2^{-k} + 13m + 5) + 4 \times 2^k - 6$$

## 6. CONCLUSIONS

In this paper, we presented the design methodologies of a novel reversible ROM. This is the first reversible design of ROM where gate count, transistor count and garbage output are optimized. For example, our proposed reversible 16×2 ROM requires 82 reversible gates, 522 transistors and produces only 18 garbage outputs. Similarly, our proposed reversible 16 × 4 ROM requires 146 reversible gates, 938 transistors and produces only 152 garbage outputs. We proposed the core components with optimum gate count, garbage outputs, quantum cost, delay and transistor count. In addition, we realized each component and related gates using MOS transistors. We also proved the efficiency of the proposed circuits by comparative results, theorems and algorithms. Simulation results showed the correctness of the proposed circuits.

Proposed reversible ROM can be used in Bootstrap memory, Embedded Microcontroller Program Memory, Firmware, Data Tables, Data Converter, Function Generator [22] etc.

### ACKNOWLEDGEMENTS

This work was done under the assistance of MOICT (Ministry of ICT) Fellowship given by the Ministry of ICT, Government of the People's Republic of Bangladesh.

### REFERENCES

[1]  R. Landauer, "Irreversibility and Heat Generation in the Computing Process", IBM journal of research and development, vol. 5, pp. 183-191, 1961.
[2]  C. H. Bennett, "Logical Reversibility of Computation", IBM journal of Research and Development, vol. 17, pp. 525-232, 1973.
[3]  V. Zhirnov, R. Cavin, J. Hutchby, G. Bourianoff, "Limits to Binary Logic Switch Scaling –a Gedanken Model", Proceedings of the IEEE, vol. 91, pp. 1934-1939, 2003.
[4]  G. E. Moore, "Cramming more components onto integrated circuits", Electronics, vol. 38, pp. 56-59, 1965.
[5]  A.S.M. Sayem, M. Ueda, "Optimization of reversible sequential circuits", Journal of Computing, vol. 38, pp. 208-214, 2010.
[6]  M.A. Nielsen, I.L. Chuang, "Quantum computation and quantum information", Cambridge university press, 2010.
[7]  R. Wille, R. Drechsler, "BDD-based synthesis of reversible logic for large functions", Proceedings of the 46th Annual Design Automation Conference, pp. 270–275. 2009.
[8]  N. Nayeem, M. Hossain, L. Jamal, H. Babu, "Efficient Design of Shift Registers Using Reversible Logic", International Conference on Signal Processing Systems, pp. 474–478, 2009.
[9]  M. Mohammadi, M.; M. Eshghi, "On figures of merit in reversible and quantum logic designs", Quantum Information Processing, vol. 8, pp. 297-318. 2009.
[10] H. Thapliyal, N. Ranganathan, "Design of Reversible Latches Optimized for Quantum Cost, Delay and Garbage Outputs", VLSID '10. 23rd International Conference on, pp. 235-240, 2010.
[11] R.P. Feynman, "Quantum mechanical computers", Foundations of physics, vol. 16, pp. 507–531, 1986.
[12] E. Fredkin, T. Toffoli, "Conservative logic", Springer, 2002.
[13] T. Toffoli, "Reversible computing", Springer, 1980.
[14] L. Jamal, M. Alam, H.M.Hasan Babu, "An efficient approach to design a reversible control unit of a processor", Sustainable Computing: Informatics and Systems, vol. 3, pp. 286-294, 2013.
[15] M. Shamsujjoha, H. Babu, "A Low Power Fault Tolerant Reversible Decoder Using MOS Transistors", International Conference on Embedded Systems and VLSI Design (VLSID), pp. 368–373, 2013.

[16] H.V.R. Aradhya, R. Chinmaye, K.N. Muralidhara, "Design, Optimization and Synthesis of Efficient Reversible Logic Binary Decoder", International Journal of Computer Applications, vol. 46, pp. 45–51, 2012.
[17] M.S.Al Mamun, I. Manda, M. Hasanuzzaman, "Design of Universal Shift Register Using Reversible Logic", International Journal of Engineering and Technology, vol. 2, pp. 1620–1625, 2012.
[18] L. Jamal, F. Sharmin, M.A.Mottalib, H.M.H. Babu, "Design and Minimization of Reversible Circuits for a Data Acquisition and Storage System", International Journal of Engineering and Technology, vol. 2, pp. 9-15, 2012.
[19] H. Thapliyal, A. Vinod, "Design of reversible sequential elements with feasibility of transistor implementation", IEEE International Symposium on Circuits and Systems, ISCAS, pp. 625–628, 2007.
[20] Y. Van Rentergem, A. De Vos, "Optimal design of a reversible full adder", International Journal of Unconventional Computing, vol. 1, pp. 339–355, 2005.
[21] DSCH: Microwind and DSCH information page.[Online].Available: http://www.microwind.org.
[22] R.J. Tocci, "Digital Systems: principles and applications", Prentice Hall, 1980.

## AUTHORS

**Sadia Nowrin** obtained her M.S. degree in Computer Science and Engineering from University of Dhaka, Bangladesh in 2015. .

**Papiya Nazneen** received a B.Sc degree in Computer Science and Engineering from University of Dhaka, Bangladesh in 2013.

**Lafifa Jamal** is currently working as an Associate Professor at the Department of Computer Science and Engineering, University of Dhaka. She received her B.Sc. and M.Sc. degrees in Computer Science from University of Dhaka, Bangladesh. She is currently pursuing her PhD in the same department. Her research interests include VLSI Design, Logic Synthesis and Design, Reversible Logic, FPGA Design and Quantum Computing. She is a member of IEEE.